# The Dark Age of the Universe

Jordi Miralda-Escudé[1,2,3]

The Dark Age is the period between the time when the cosmic microwave background was emitted and the time when the evolution of structure in the universe led to the gravitational collapse of objects, in which the first stars were formed. The period of reionization started with the ionizing light from the first stars, and it ended when all the atoms in the intergalactic medium had been reionized. The most distant sources of light known at present are galaxies and quasars at redshift $z \cong 6$, and their spectra indicate that the end of reionization was occurring just at that time. The Cold Dark Matter theory for structure formation predicts that the first sources formed much earlier.

It was only about 75 years ago when Edwin Hubble discovered that we live in a universe of galaxies in expansion. At about the same time, Alexander Friedmann used the cosmological principle (the assumption that the universe can be approximated on large scales as homogeneous and isotropic) to write down the basic equations governing the structure and evolution of the universe in the Big Bang model, starting from Einstein's theory of General Relativity. By the end of the 20th century, much evidence had accumulated showing that the early universe was close to homogeneous, even on the small scales of the present galaxies. The fundamental question is how the universe went from this initial nearly homogeneous state to the present-day extremely complex form, in which matter has collapsed into galaxies and smaller structures.

I will review the history of the universe from the time of emission of the cosmic microwave background (CMB) to the time when the first objects collapsed gravitationally. An overview of these events will be described, with respect to the time and the redshift at which they take place (Fig. 1). Cosmologists generally use the redshift $z$ to designate a cosmic epoch. The quantity $1 + z$ is the factor by which the universe has expanded from that epoch to the present time and is also the factor by which the wavelength of the light emitted by any object at that epoch and reaching us at the present time has been stretched, owing to the expansion of the universe.

## The Cold Dark Matter Model

Cosmological observations can be accounted for by the Cold Dark Matter (CDM) model [see (1–3) for reviews]. The model assumes that in addition to ordinary matter made of protons, neutrons, and electrons (usually re-geneous and in thermal equilibrium) also reveal that for these primordial, small-amplitude fluctuations to have grown into the present galaxies, clusters, and large-scale structures of the universe through gravitational evolution, the presence of dark matter is required. More recently, another component has been identified, called dark energy, which has become the dominant component of the universe at the present epoch and is causing an acceleration of the expansion of the universe (10, 11).The Wilkinson Microwave Anisotropy Probe (WMAP) (12, 13) showed that the baryonic matter accounts for only ~17% of all matter, with the rest being the dark matter, and has confirmed the presence of the dark energy (14, 15). Although the CDM model with the added dark energy agrees with many observations, cosmologists have no idea what the nature of the dark matter and ferred to as baryonic matter in cosmology), there is also dark matter, which behaves as a collection of collisionless particles having no interactions other than gravity and which was initially cold (that is, the particles had a very small velocity dispersion). Observations have confirmed the existence of dark matter in galaxy halos and clusters of galaxies [e.g., (4–9)]. The intensity fluctuations of the CMB (the relic radiation that is left over from the epoch when the universe was nearly homo-the dark energy may be, and why this matter and energy should have comparable densities at the present time.

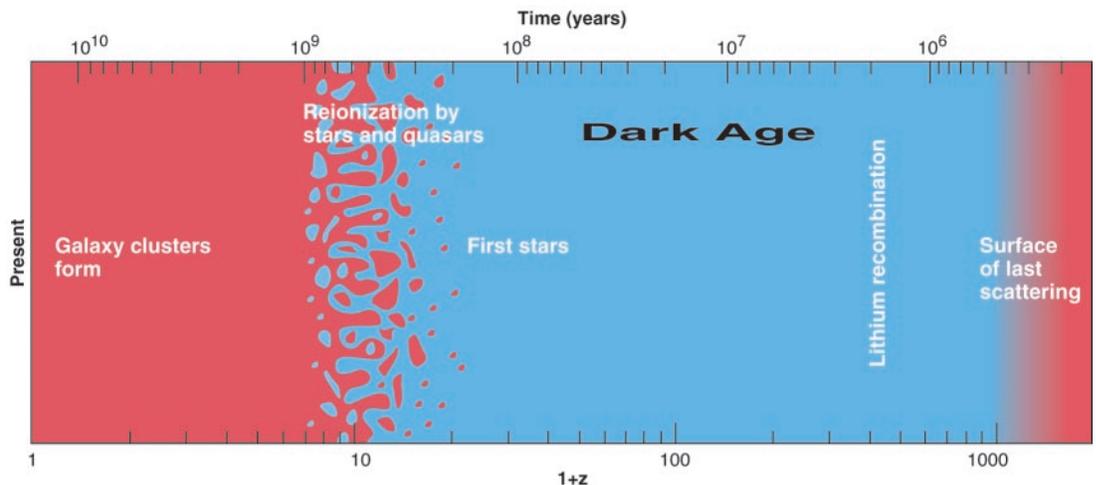

**Fig. 1.** Overview of the main events discussed in this review, with the top axis showing the age of the universe and the bottom axis the corresponding redshift, for the currently favored model (same parameters as in Fig. 2). Blue represents atomic regions, and red, ionized regions. Matter in the universe recombined in a homogeneous manner at $z \cong 1200$. Later, when the first stars formed and emitted ionizing radiation, ionized regions formed around the sources that eventually overlapped, filling all of space. The size of the HII regions should be much smaller on the redshift scale than shown here and is drawn only for illustration.

Nevertheless, as the parameters of this CDM model are measured more precisely, the predictions for the number of objects of different mass that should be gravitationally collapsing at every epoch in the universe have become more robust. Bound objects form when the primordial fluctuations reach an amplitude near unity, entering the nonlin-ear regime. The power spectrum of the fluctuations can be represented in terms of the root-mean-square (rms) fluctuation of the mass, $\delta M$, enclosed by a sphere of radius $R$, which on average has a mass $M$, equal to its volume times the mean density of the universe. The linearly extrapolated rms fluctuation $\delta M/M$ is shown in Fig. 2 as a function of $M$ for the CDM model, at the present time ($z = 0$) and at redshifts $1 + z = 10$ and $1 + z = 20$. Note that linear fluctuations grow gravitationally in proportion to $(1 + z)^{-1}$, except at $z \lesssim 1$, when the dark energy starts to dominate (16). At the present time, fluctuations are typically of order unity on scales containing masses $\sim 10^{14} \, M_\odot$ (where $M_\odot$ is solar mass), corresponding to galaxy groups. At the epoch $z = 9$, typical fluctuations were collapsing on much smaller scales of $M \sim 10^6 \, M_\odot$. Because the probability distribution of the mass fluctuation on any given region is Gaussian, there should be rare regions in the universe with a density fluctuation of several times the variance that will correspondingly be able to collapse earlier. For example, our Milky Way galaxy may have formed from the collapse of a $10^{12} \, M_\odot$ halo from a 1σ fluctuation at $z = 1$, but at $z = 5$ halos of the same mass were already forming from 3σ fluctuations. On a scale of $10^6 M_\odot$, a 1σ fluctuation collapses at $z = 6$, and a 3σ fluctuation collapses at $z = 20$ (Fig. 3). Each object that forms has a velocity dispersion $v$ determined by its mass and the size of the region from which it collapsed, $v^2 \sim GM/R$, and a corresponding virialized temperature of the gas, $kT_{vir} = (\mu m_H) \, v^2$, where $\mu$ is the mean particle mass in units of the hydrogen mass $m_H$. It is this virialized temperature that determines the physics of the rate at which gas can cool to form stars. This prediction of the number of objects that were forming at each $z$ forms the basis for our ideas on the end of the Dark Age, the formation of the first stars, and the reionization.

[1]Department of Astronomy, The Ohio State University, Columbus, OH 43210, USA. [2]Institute for Advanced Study, Princeton, NJ 08540, USA. [3]Institut d'Estudis Espacials de Catalunya/ICREA, Barcelona, Spain. E-mail: jordi@astronomy.ohio-state.edu

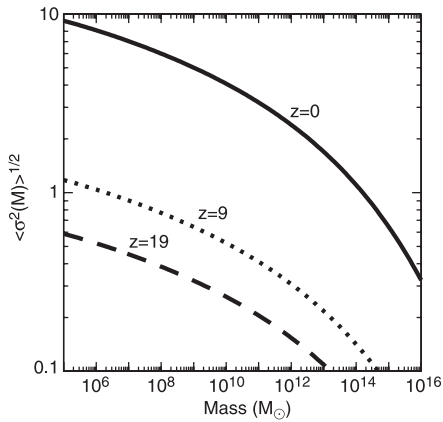

**Fig. 2.** The solid line shows the present time ($z = 0$), linearly extrapolated rms fluctuation $<(\delta M/M)^2>^{1/2}$ of the mass enclosed in a region that contains an average mass $M$, expressed in the horizontal axis in units of solar masses. The other two curves are for $z = 9$ and $z = 19$, when the universe was about 500 million and 200 million years old, respectively. Fluctuations grow with time, and when they reach an amplitude near unity at some scale, nonlinear formation of halos takes place, and small halos merge into larger ones as progressively larger scales undergo collapse. The flat CDM model with cosmological constant assumed here has the following parameters: Hubble constant $H_0 = 70$ km s$^{-1}$ Mpc$^{-1}$, matter density $\Omega_{m0} = 0.3$, baryon density $\Omega_b = 0.043$, amplitude of fluctuations $\sigma_8 = 0.9$, and primordial spectral index $n = 0.93$.

### The Dark Age

At very high $z$, the universe was practically homogeneous, and the temperature of matter and radiation dropped as the universe expanded. Atoms formed at $z \cong 1100$ when the temperature was $T = 3000$ K, a low enough value for the plasma to recombine. At this epoch of recombination, the CMB filled the universe with a red, uniformly bright glow of blackbody radiation, but later the temperature dropped and the CMB shifted to the infrared. To human eyes, the universe would then have appeared as a completely dark room. A long period of time had to pass until the first objects collapsed, forming the first stars that shine in the universe with the first light ever emitted that was not part of the CMB (Fig. 1). The period of time between the last scattering of the CMB radiation by the homogeneous plasma and the formation of the first star has come to be known as the Dark Age of the universe (17).

Observations provide detailed information on the state of the universe when the CMB radiation was last scattered at $z \cong 1100$, and we have also observed galaxies and quasars up to $z \cong 6.5$ (18–21). The theory suggests that the first stars and galaxies should have formed substantially earlier, so we can expect to discover galaxies at progressively higher $z$ as technology advances and fainter objects are detected. However, beyond a $z$ of 10 to 20, the CDM theory with Gaussian fluctuations predicts that the dark matter halos that can host luminous objects become extremely rare, even for low-mass halos (Fig. 2). Discovering any objects at $z \gtrsim 20$ will become exceedingly difficult as we reach the period of the Dark Age. During the Dark Age, before the collapse of any objects, not much was happening at all. The atomic gas was still close to homogeneous, and only a tiny fraction of it formed the first molecules of $H_2$, HD, and LiH as the temperature cooled down [e.g., (22, 23)]. One of the few suggested ideas for an observational probe of the Dark ge is to detect secondary anisotropies on the CMB that were imprinted by Li atoms as they recombined at $z \cong 400$ through the resonance line at 670.8 nm, which would be redshifted to the far-infrared today, making it difficult to observe because of the foreground emission by dust (24, 25).

### How Did the First Stars Form?

The Dark Age ended when the first stars were formed. In order to form stars, the atomic gas must be able to follow the collapse of dark matter halos. This happens when the halo mass is above the Jeans mass of the gas (26) at the virialized temperature and density of the intergalactic medium, a condition that is fulfilled when $T_{vir} \gtrsim 100$ K (1, 27). In halos with lower temperature, the gas pressure is sufficient to prevent the gas from collapsing. In addition, there must be a radiative cooling mechanism for the gas to lose its energy and concentrate to ever-higher densities in the halo centers until stellar densities are reached; without cooling, the gas reaches hydrostatic equilibrium in the halo after the gravitational collapse and stays at a fixed density without forming stars. The ability of the gas to cool depends on $T_{vir}$ and the chemical composition of the gas. $T_{vir}$ was low for the first objects that formed and then it increased rapidly with time (Fig. 3). The primordial gas in the first halos was mainly composed of atomic H and He. Atomic H induces radiative cooling only when $T_{vir} > 10^4$ K, when collisions can excite and ionize H atoms (28); the gas can then readily con-tract to form galaxies. In the intermediate range $100 \text{ K} < T_{vir} < 10^4 \text{ K}$, the gas settles into halos but atomic cooling is not available and, in the absence of the heavy elements that were formed only after massive stars ejected their synthesized nuclei into space, the only available coolant is $H_2$. Because two hydrogen atoms cannot form a molecule by colliding and emitting a photon, only a small fraction of the gas in these first objects could become $H_2$ via reactions involving the species $H^-$ and $H_2^+$, formed by the residual free electrons and protons left over from the early universe (29–31), limiting the rate at which the gas could cool. Simulations (32–37) have shown that the first stars form in halos with $T_{vir} \approx 2000 \text{ K}$ and mass $\sim 10^6 \, M_\odot$; at lower temperatures, the rotational transitions of $H_2$ do not provide sufficient cooling for the gas to dissipate its energy. The slow cooling in these first objects leads to the formation of a central core with a mass of 100 to 1000 $M_\odot$ of gas cooled to $\sim 200 \text{ K}$, and this core may form a massive star.

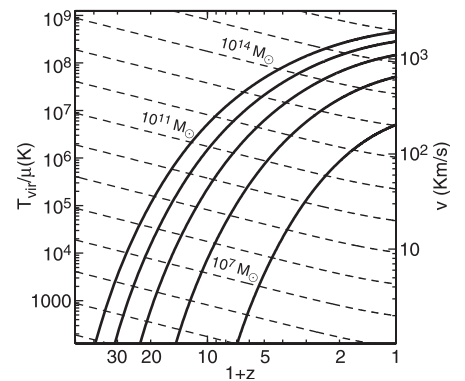

**Fig. 3.** The velocity dispersion $v$ (right axis) or virialized temperature $T_{vir}$ divided by the mean particle mass $\mu$ in units of the hydrogen mass (left axis; $\mu = 0.6$ for ionized matter and $\mu = 1.2$ for atomic matter) of halos collapsing from a 1σ fluctuation (of amplitude shown in Fig. 2) is shown as a function of redshift, as the lowest thick solid line. At every redshift, the fluctuation amplitude required for nonlinear collapse is reached at progressively larger scales, forming halos of increasing mass and velocity dispersion. The higher solid thick lines indicate halos collapsing from (2,3,4,5)-σ fluctuations, which form increasingly rare objects from a Gaussian distribution of fluctuations. The dashed lines indicate halos of constant mass, and are separated by a factor 10 in mass, with values indicated for three lines. Objects of fixed mass have increasing velocity dispersion as they form at higher redshift from a more rare, higher amplitude fluctuation because their size $R$ is smaller.

As soon as the first stars appeared, they changed the environment in which they were formed, affecting the formation of subsequent stars. Massive stars emit a large fraction of their light as photons that can ionize H (with energies greater than 13.6 eV), creating HII regions and heating the gas to $T \approx 10^4 \text{ K}$. While these ionizing photons are all absorbed at the HII region boundaries, in the vicinity of the stars that emit them, photons with lower energy can travel greater distances through the atomic medium and reach other halos. Ultraviolet photons with energies above 11 eV can photodissociate $H_2$, and this can suppress the cooling rate and the ability to form stars in low-mass halos that are cooling by $H_2$ when they are illuminated by the first stars (38). The importance of this suppresion and other effects are being debated (37, 39–43). Such effects might imply that the first massive stars formed through the radiative cooling of $H_2$ were a short-lived and self-destructive generation, because their own light might destroy the molecules that made their formation possible.

When some of these massive stars end their lives in supernovae, they eject heavy elements that pollute the universe with the ingredients necessary to form dust and planets (44). In a halo containing $10^6 \, M_\odot$ of gas, the photoionization and supernova explosions from only a few massive stars can expel all the gas from the potential well of the halo (45). For example, the energy of 10 supernovae (about $10^{52}$ erg) is enough to accelerate $10^6 \, M_\odot$ of gas to a speed of 30 km s$^{-1}$, which will push the gas out of any halo with a much lower velocity dispersion. The expelled gas can later fall back as a more massive object is formed by mergers of pre-existing dark matter halos. The next generation of stars can form by cooling provided by heavy elements (46), or by atomic H when $T_{vir} > 10^4 \text{ K}$. Abundances of heavy elements as low as 1000 times smaller than that of the sun can increase the cooling rate over that provided by $H_2$ and can also cool the gas to much lower temperatures than possible with $H_2$ alone, reducing the Jeans mass and allowing for the formation of low-mass stars (47–49).

A fascinating probe to these early events is provided by any stars that formed at that time with mass $\sim 0.8$ solar masses, which could be observed at the present time in our Galaxy's halo as they start ascending the red giant branch (50) if the halos in which they formed were later incorporated into the Milky Way by mergers. These stars should carry the signature of the elements synthesized by the first supernova (51, 52).

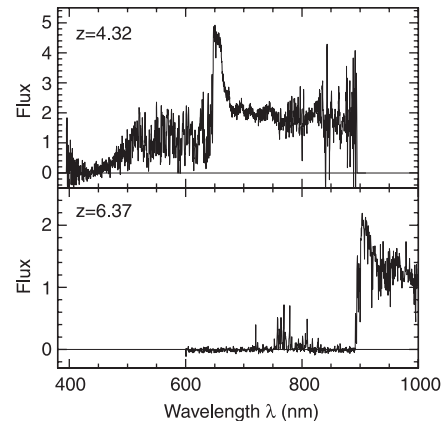

**Fig. 4.** Spectra of the Sloan Digital Sky Survey quasars J0019-0040 at $z = 4.32$, and J1148+5251 at $z = 6.37$. The flux is shown in units of $10^{-17}$ erg cm$^{-2}$ s$^{-1}$ as a function of wavelength. The peak of the spectra is the redshifted broad Ly$\alpha$ emission line of the quasars. Absorption by intervening hydrogen is seen at shorter wavelengths. At redshifts below 6 ($\lambda \lesssim 850$ nm), the medium is photoionized and the very small fraction of hydrogen that is atomic produces a partial, strongly fluctuating absorption reflecting the density variations of the intergalactic medium. At $z \approx 6$, the absorption suddenly becomes complete. This probably indicates the end of reionization. At $z > 6$, the medium still contained atomic patches that are highly opaque to Ly$\alpha$ photons, and, even in the reionized regions, the ionizing background intensity was too low to reduce the neutral fraction to the very low values required for Ly$\alpha$ transmission. This figure is reproduced from [(18) fig. 3] and [(9) fig. 6].

### When Did the First Star Form?

Because the primordial density fluctuations in the universe are random, the question of when the very first star formed does not have a simple answer. The time when the first halo with $T_{vir} = 2000 \text{ K}$ collapsed depends on how rare a fluctuation we are willing to consider. A 5σ fluctuation in the density field can lead to the collapse of a halo and the formation of a star at $z \approx 30$ (Fig. 3). A more specific question we can ask is: From a random location in the universe, when would the first light from a star have been observed? Because an observer receives light only from the past light-cone (53), the further away one looks, the greater the volume that can be surveyed (and hence a more rare, higher-amplitude fluctuation can be found) but also the further back into the past one observes, which requires an even higher primordial density fluctuation to form a star. By requiring that just one collapsed halo with $T_{vir} > 2000 \text{ K}$ is observed on the past light-cone [and for the CDM model (Fig. 3)], a hypothetical observer located at a random place, after having experienced the dark age, would have seen the first star appear in the sky at $z \approx 38$ (54), when the universe was 75 million years old. This star would have formed from a 6.3σ fluctuation (with a probability of only $\sqrt{2/\pi} \int_{6.3}^{\infty} e^{-x^2/2} \, dx \approx 3 \times 10^{-10}$, implying that a volume containing a mass of $10^6 \, M_\odot / 3 \times 10^{-10} \approx 3 \times 10^{15} \, M_\odot$ would need to be searched to find one halo of $10^6 M_\odot$ at this early time). Soon after that first star, many more would have appeared forming from less rare fluctuations.

Because we can now see the very first stars that formed in the universe out to a very large distance on our past light-cone, we can survey a much larger volume than could the overjoyed observer at $z \approx 38$ at the sight of the first star. With this larger volume, the highest $z$ star on the sky should be one formed from an 8σ fluctuation at $z \approx 48$ (54). Although this first star would be too faint to detect with current technology, brighter sources can pave the way to discover more primitive objects than the presently known most distant galaxies at $z \approx 6.5$. Perhaps we may discover more objects at higher $z$ than expected in the CDM model, for example due to the presence of non-Gaussian primordial fluctuations on small scales [e.g., (55)].

### The Reionization of the Universe

The most important effect that the formation of stars had on their environment was the reionization of the gas in the universe. Even though the baryonic matter combined into atoms at $z \approx 1100$, the intergalactic matter must have been reionized before the present. The evidence comes from observations of the spectra of quasars. Quasars are extremely luminous objects found in the nuclei of galaxies that are powered by the accretion of matter on massive black holes (56). Because of their high luminosity, they are used by cosmologists as lamp posts allowing accurate spectra to be obtained, in which the analysis of absorption lines provides information on the state of the intervening intergalactic matter. The spectra of quasars show the presence of light at wavelengths shorter than the Lyman-alpha (Ly$\alpha$) emission line of H. If the intergalactic medium is atomic, then any photons emitted at wavelengths shorter than Ly$\alpha$ (121.6 nm) would be scattered by H at some point on their journey to us, when their wavelength is redshifted to the Ly$\alpha$ line. The mean density of H in the universe, when it is all in atomic form, is enough to provide a scattering optical depth as large as $\sim 10^5$ [57]. The suppression of the flux at wavelengths shorter than the Ly$\alpha$ emission line is called the Gunn-Peterson trough.

In quasars at $z < 6$, the Gunn-Peterson trough is not observed. Instead, one sees the flux partially absorbed by what is known as the Ly$\alpha$ forest: a large number of absorption lines of different strength along the spectrum (Fig. 4). The H atoms in the intergalactic medium producing this absorption are a small fraction of all of the H, which is in photoionization equilibrium with a cosmic ionizing background produced by galaxies and quasars [58]. The absorption lines correspond to variations in the density of the intergalactic matter. The observation that a measurable fraction of Ly$\alpha$ flux is transmitted through the universe implies that, after $z = 6$, the entire universe had been reionized.

However, recently discovered quasars [19, 59, 60] show a complete Gunn-Peterson trough starting at $z \cong 6$ (Fig. 4). Although the lack of transmission does not automatically imply that the intervening medium is atomic (because the optical depth of the atomic medium at mean density is $\sim 10^5$, and so even an atomic fraction as low as $10^{-3}$ produces an optical depth of $\sim 100$, which implies an undetectable transmission fraction), analysis of the Ly$\alpha$ spectra in quasars at $z < 6$ [61, 62] indicates that the intensity of the cosmic ionizing background increased abruptly at $z \cong 6$. The reason for the increase has to do with the way in which reionization occurred. Ionizing photons in the far-ultraviolet have a short mean free path through atomic gas in the universe, so they are generally absorbed as soon as they reach any region in which the gas is mostly atomic. Initially, when the first stars and quasars were formed, the ionizing photons they emitted were absorbed in the high-density gas of the halos hosting the sources. The intergalactic medium started to be reionized when sufficiently powerful sources could ionize all the gas in their own halos, allowing ionizing photons to escape. The reionization then proceeded by the expansion of ionization fronts around the sources (Fig. 5), separating the universe into ionized bubbles and an atomic medium between the bubbles [63]. The ionized bubbles grew and overlapped, until every low-density region of the universe was reionized; this moment defines the end of the reionization period. High-density regions that do not contain a luminous internal source can remain atomic because the gas in them recombines sufficiently fast, and they can self-shield against the external radiation. When the ionized bubbles overlap, photons are free to travel for distances much larger than the size of a bubble before being absorbed, and the increase in the mean free path implies a similar increase in the background intensity. The exact way in which the background intensity should increase at the end of reionization, depending on the luminosity function and spatial distribution of the sources, has not yet been predicted by theoretical models of reionization [e.g., [64]], but a rapid increase in the mean free path should, if present, tell us the time at which the reionization of the low-density intergalactic medium was completed.

The observational pursuit of the reionization epoch may be helped by the optical afterglows of gamma-ray bursts, which can shine for a few minutes with a flux that is larger than even the most luminous quasars [65–70], probably due to beaming of the radiation. Because gamma-ray bursts may be produced by the death of a massive star, they can occur even in the lowest-mass halos forming at the earliest times, with fixed luminosities. Among other things, the absorption spectra of gamma-ray burst optical afterglows might reveal the damped Ly$\alpha$ absorption profile of the H in the intervening atomic medium [68] and absorption lines produced by neutral oxygen (which can be present in the atomic medium only, before reionization) ejected by massive stars [71, 72].

## Electron Scattering of the CMB by the Reionized Universe

Reionization made most of the electrons in the universe free of their atomic binding, and able to scatter the CMB photons again. Before recombination at $z = 1100$, the universe was opaque, but because of the large factor by which the universe expanded from recombination to the reionization epoch, the electron Thompson scattering optical depth produced by the intergalactic medium after reionization, $\tau_e$, is low. If the universe had reionized suddenly at $z = 6$, the $\tau_e \cong 0.03$. Because the fraction of matter that is ionized must increase gradually, from the time the first stars were formed to the end of reionization at $z = 6$, $\tau_e$ must include the contribution from the partially ionized medium at $z > 6$, and it must therefore be greater than 0.03.

The sooner reionization started, the larger the value of $\tau_e$.

The WMAP mission has measured $\tau_e$ from the power spectrum of the polarization and temperature fluctuations of the CMB. A model-independent measurement from the polarization-temperature correlation gives $\tau_e = 0.16 \pm 0.04$ [73], but a fit to the CDM model with six free parameters using both the correlation of temperature and polarization fluctuations found by WMAP, and other data gives $\tau_e = 0.17 \pm 0.06$ [13]. An optical depth as large as $\tau_e = 0.16$ is surprising because it implies that a large fraction of the matter in the universe was reionized as early as $z = 17$, when halos with mass as low as $10^7 M_\odot$ could collapse only from $3\sigma$ peaks, and were therefore still very rare (Fig. 3). The errors on $\tau_e$ will need to be reduced before we can assign a high degree of confidence to its high value [74].

What are the implications of a high $\tau_e$ if it is confirmed? Measurements of the emission rate at $z \cong 4$ from the Ly$\alpha$ forest show that to obtain $\tau_e > 0.1$, the emission rate would need to increase with $z$ [75], and a large increase is required up to $z = 17$ to reach $\tau_e = 0.16$. In view of the smaller mass fraction in collapsed halos at this high $z$, it is clear that a large increase in the ionizing radiation emitted per unit mass is required from $z = 6$ to 17. Models have been proposed to account for an early reionization, based on a high emission efficiency at high $z$ (76–84). A possible reason for this high efficiency is that if the first stars that formed with no heavy elements were all massive (34, 36), they would have emitted as many as $10^5$ ionizing photons per baryon in stars (85), many more than emitted by observed stellar populations (86–89). It is not clear, however, if enough of these massive stars can form in the first low-mass halos at $z > 17$, once the feedback effects of ultraviolet emission and supernovae (37, 38, 45) are taken into account. A different possibility might be that more objects than expected were forming at high z due to a fundamental change in the now favored CDM model.

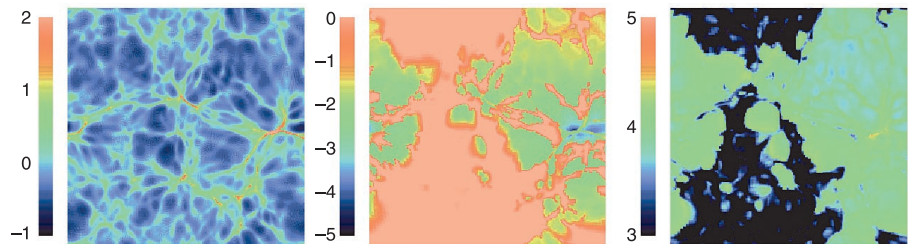

**Fig. 5.** Results of a simulation of the reionization of the intergalactic medium in a cubic box of co-moving side $4 \ h^{-1}$ Mpc, from [64] (Fig. 3B). The gas density (**left panel**), neutral fraction (**central panel**), and temperature (**right panel**) from a slice of the simulation are shown. The color coded values indicate the logarithms of the gas density divided by the mean baryon density, the neutral fraction, and the gas temperature in Kelvin, respectively. The simulation is shown at $z = 9$. The pink regions in the central panel are atomic, and the green regions are ionized. The sources of ionizing photons generally appear in halo centers where the gas density is high, but once the photons escape from the local high-density regions, the ionized bubbles expand most easily across the lowest density regions (compare left and central panels). The ionized regions are heated to about $10^4$ K (see right panel), and they grow with time until they fill the entire universe at the end of reionization.

## The Future: the 21-cm Signature of the Atomic Medium

Many of the observational signatures of the epoch of reionization probe regions of the universe where stars have already formed and the medium has been reionized or polluted by heavy elements. But there is a way to study the undisturbed atomic medium. Nature turns out to be surprisingly resourceful in providing us with opportunities to scrutinize the most remote landscapes of the universe. The hyperfine structure of H atoms, the 21-cm transition due to the spin interaction of the electron and the proton, provides a mechanism to probe the atomic medium. When observing the CMB radiation, the intervening H can change the intensity at the redshifted 21-cm wavelength by a small amount, causing absorption if its spin temperature is lower than the CMB temperature, and emission if the spin temperature is higher. The spin temperature reflects the fraction of atoms in the ground and the excited hyperfine levels. The gas kinetic temperature cooled below the CMB temperature during the Dark Age owing to adiabatic expansion, although the spin temperature was kept close to the CMB temperature (90). When the first stars appeared in the universe, a mechanism for coupling the spin and kinetic temperature of the gas, and hence for lowering the spin temperature and making the H visible in absorption against the CMB, started to operate. The ultraviolet photons emitted by stars that penetrated the atomic medium were repeatedly scattered by H atoms after being redshifted to the Lyα resonance line, and these scatterings redistributed the occupation of the hyperfine structure levels (90–93), bringing the spin temperature down to the kinetic temperature and causing absorption. As the first generation of stars evolved, supernova remnants and x-ray binaries probably emitted x-rays that penetrated into the intergalactic medium and heated it by photoionization; gas at high density could also be shock-heated when collapsing into halos. The gas kinetic and spin temperatures could then be raised above the CMB, making the 21-cm signal observable in emission (93, 94). This 21-cm signal should reveal an intricate angular and frequency structure reflecting the density and spin temperature variations in the atomic medium (95–100). Several radio observatories will be attempting to detect the signal (101).

The observation of the 21-cm signal on the CMB will be a challenge, because of the long wavelength and the faintness of the signal. However the potential for the future is enormous: detailed information on the state of density fluctuations of the atomic medium at the epoch when the first stars were forming and the spin temperature variations that were induced by the ultraviolet and x-ray emission from the first sources are both encoded in the fine ripples of the CMB at its longest wavelengths.

REVIEW

# New Light on Dark Matter

Jeremiah P. Ostriker[1] and Paul Steinhardt[2]

Dark matter, proposed decades ago as a speculative component of the universe, is now known to be the vital ingredient in the cosmos: six times more abundant than ordinary matter, one-quarter of the total energy density, and the component that has controlled the growth of structure in the universe. Its nature remains a mystery, but assuming that it is composed of weakly interacting subatomic particles, is consistent with large-scale cosmic structure. However, recent analyses of structure on galactic and subgalactic scales have stimulated discrepancies and stimulated numerous alternative proposals. We discuss how studies of the density, demography, history, and environment of smaller-scale structures may distinguish among these possibilities and shed new light on the nature of dark matter.

The dark side of the universe first became evident about 65 years ago when Fritz Zwicky (1) noticed that the speed of galaxies in large clusters is much too great to keep them gravitationally bound together unless they weigh over 100 times more than one would estimate on the basis of the number of stars in the cluster. Decades of investigation confirmed his analysis (2–5), but by the 1980s, the evidence for dark matter with an abundance of about 20% of the total energy density of the universe was accepted, although the nature of the dark matter remained a mystery.

After the introduction of inflationary theory (6), many cosmologists became convinced that the universe must be flat and that the total energy density must equal the value (termed the critical value) that distinguishes a positively curved, closed universe from a negatively curved, open universe. Cosmologists became attracted to the beguiling simplicity of a universe in which virtually all of the energy density consists of some form of matter, about 4% being ordinary matter and 96% dark matter. In fact, observational studies were never really compliant with this vision. Although there was a wide dispersion in total mass density estimates, there never developed any convincing evidence that there was sufficient matter to reach the critical value. The discrepancy between observation and the favored theoretical model became increasingly sharp.

Dark energy came to the rescue when it was realized that there was not sufficient matter to explain the structure and nature of the universe (7). The only thing dark energy has in common with dark matter is that both components neither emit nor absorb light. On a microscopic scale, they are composed of different constituents. Most important, dark matter, like ordinary matter, is gravitationally self-attractive and clusters with ordinary matter to form galaxies. Dark energy is gravitationally self-repulsive and remains nearly uniformly spread throughout the universe. Hence, a census of the energy contained in galaxies would miss most of the dark energy. So, by positing the existence of a dark energy component, it became possible to account for the 70 to 80% discrepancy between the measured mass density and the critical energy density predicted by inflation (8–11). Then, two independent groups (12, 13) found evidence of the accelerated expansion of the universe from observations of supernovae, and the model with a dominant dark energy component, as illustrated in Fig. 1, became the concordance model of cosmology. The existence of dark energy has recently been independently confirmed by observations by the Wilkinson Microwave Anisotrope Probe [WMAP (14)] and has become accepted as an essential ingredient of the standard model (15).

Dark energy has changed our view of the role of dark matter in the universe. According to

[1]Department of Astrophysical Sciences, [2]Department of Physics, Princeton University, Princeton, NJ 08544, USA.